\documentclass[a4paper,twocolumn,english,superscriptaddress]{revtex4}
\usepackage{mathptmx}
\usepackage{helvet}
\usepackage{courier}
\usepackage[T1]{fontenc}
\usepackage[latin9]{inputenc}
\usepackage{color}
\usepackage{textcomp}
\usepackage{relsize}
\usepackage{amsmath}
\usepackage{graphicx}
\usepackage{amssymb}

\makeatletter

\DeclareRobustCommand{\lyxmathsym}[1]{\ifmmode\begingroup\def\b@ld{bold}
  \def\rmorbf##1{\ifx\math@version\b@ld\textbf{##1}\else\textrm{##1}\fi}
  \mathchoice{\hbox{\rmorbf{#1}}}{\hbox{\rmorbf{#1}}}
  {\hbox{\smaller[2]\rmorbf{#1}}}{\hbox{\smaller[3]\rmorbf{#1}}}
  \endgroup\else#1\fi}

\@ifundefined{textmu}
 {\usepackage{textcomp}}{}

\makeatother

\usepackage{babel}

\begin{document}

\title{Depth dependent local structures in thin films unraveled by grazing
incidence x-ray absorption spectroscopy}

\author{Narcizo M. Souza-Neto}

\altaffiliation{Present address: Advanced Photon Source, Argonne National Laboratory, Argonne, Illinois 60439, USA}

\affiliation{Laboratório Nacional de Luz Síncrotron, CP 6192, 13083-970, Campinas,
Brazil}

\affiliation{Departamento de Física dos Materiais e Mecânica, Instituto de Física,
Universidade de São Paulo, São Paulo, Brazil}

\author{Aline Y. Ramos}

\affiliation{Institut Néel, CNRS et Université Joseph Fourier, BP 166, F-38042
Grenoble Cedex 9, France}

\affiliation{Laboratório Nacional de Luz Síncrotron, CP 6192, 13083-970, Campinas,
Brazil}

\author{Hélio C. N. Tolentino}

\affiliation{Institut Néel, CNRS et Université Joseph Fourier, BP 166, F-38042
Grenoble Cedex 9, France}

\affiliation{Laboratório Nacional de Luz Síncrotron, CP 6192, 13083-970, Campinas,
Brazil}

\author{Alessandro Martins}

\affiliation{Universidade Federal de Goiás, Campus Jataí, Jataí, Brazil}

\author{Antonio D. Santos}

\affiliation{Departamento de Física dos Materiais e Mecânica, Instituto de Física,
Universidade de São Paulo, São Paulo, Brazil}
\begin{abstract}
A method of using X-ray absorption spectroscopy (XAS) together with
resolved grazing incidence geometry for depth profiling atomic, electronic,
chemical or magnetic local structures in thin films is presented.
The quantitative deconvolution of thickness-dependent spectral features
is performed by fully considering both scattering and absorption formalisms.
Surface oxidation and local structural depth profiles in nanometric
FePt films are determined, exemplifying the application of the method. 
\end{abstract}
\maketitle

\section{Introduction}

Magnetic thin films have attracted a lot of attention due to their
extremely high-density magnetic recording applications~\citep{Weller-ARMS00}.
In this regard, a clear understanding of the macroscopic magnetic
properties requires a substantial knowledge of its dependence with
layers thicknesses \citep{Johnson-RPP96} and the complex microstructural
effects frequently localized at the interface with the substrate or
the surface of the films. Such effects can be studied using experimental
techniques able to peer selectively in the depth of the films. In
a previous letter \citep{Souza-Neto-APL06} we presented qualitative
results using x-ray absorption spectroscopy (XAS) with resolved grazing
incidence (GI) to clarify the thickness-dependent magnetic properties
in nanometric CoPt films. A depth dependent chemical order was revealed
and the magnetic behavior was interpreted within this framework. In
the present paper we provide a rigorous quantitative method for the
deconvolution of the local atomic, chemical and magnetic structural
depth profiles. This method is then illustrated by \textcolor{black}{its}
application to oxidized FePt thin films~\citep{Martins-JMMM06}.
The proposed approach makes GI-XAS a unique tool to address the depth
dependence of the local structural parameters, suitable for nanometric
structures where this dependence is a crucial issue. Moreover this
method provides a new venue to rigorously determine depth dependent
electronic structure profiles using XANES (x-ray absorption near edge
structure), which turns out to be crucial in understanding striking
artifical interface materials \citep{Souza-Neto-PRL09,Freeland-Science07}.

Although the general phenomena of scattering and absorption of x-rays
by the condensed matter are nowadays quite well understood \citep{Nielsen-01},
they still are normally explored from unconnected viewpoints. A few
well established techniques surpass this general rule with interconnected
scattering and absorption techniques, providing invaluable additional
selectivity compared to each approach used separately. DAFS (diffraction
anomalous fine structure) gives site selectivity and local structural
information \citep{Stragier-PRL92}. XAFS (x-ray absorption fine structure)
extracted from reflectivity data gives local structural information
from surfaces and interfaces \citep{Keil-EPL05}. XSW (x-ray standing
wave) locates impurities in bulk crystals and nanostructures using
an interference field that provides spatial dependence to the x-ray
spectroscopic yields from atoms within the field \citep{Bedzyk-XSW}.
Similarly, glacing-incidence x-ray fluorescence (GIXRF) is a sensitive
probe of chemical composition as a function of depth \citep{DeBoer-PRB91}.
These techniques are based on similar approaches to the one presented
here, however they are limited to near-perfect crystal structures,
require well defined geometries and/or give a limited set of information.

\section{Grazing incidence x-ray absorption spectroscopy}

X-ray absorption spectra contain information about the ground state
of the selected element in a material (local symmetry, oxidation and
spin states, spin-orbit coupling in the $2p$ and $3d$ orbitals,
crystal field, covalence and charge transfer). As a matter of fact,
in the case of $3d$ transition metals essentially structural information
is obtained from the K edges, while more magnetic and electronic information
is usually deduced from $\mathrm{L}_{2,3}$ edge. XAS is not a surface
technique by itself, since the atenuation length of hard x-rays is
of a few micrometers in any material. However, in the grazing incidence
geometry near the critical angle for total reflexion, the x-ray beam
is confined within a few nanometers from the surface. For this film
studies, this confinement has the considerable advantage of minimizing
the substrate contribution. 

The grazing incidence x-ray absorption measurements were performed
at the Brazilian Synchrotron Light Laboratory (LNLS - Laboratório
Nacional de Luz Síncrotron). The setup includes 20 $\mu$m-vertical
slits limiting the beam size on the sample mounted on a high precision
goniometer. XANES spectra were collected in the fluorescence mode
at the D04B-XAFS1 beamline \citep{Tolentino-JSR01} with a Si (111)
channel-cut monochromator. The incident beam intensity was monitored
using a first ion-chamber. The reflected beam and fluorescence emission
were simultaneously collected using a second ion-chamber and a 15-elements
Ge detector, respectively. The fluorescence emission and/or x-ray
reflectivity curves were used to calibrate and select with an accuracy
of $\approx0.01^{o}$ the working grazing angle corresponding to a
chosen penetration depth profile. For an accurate energy calibration,
the transmission through \textcolor{black}{an Iron} metal reference
foil was monitored using a third ion-chamber.

The collected absorption spectra measured by the fluorescence yield
is a mix of contributions coming from different depths. To get quantitative
information we must deconvolve them into their absorption contributions
from each depth (z) into the films at each photon energy (E) and grazing
angle ($\theta$). The electromagnetic radiation amplitude at each
set of (E, $\theta$, z) must be known to weightly sum the absorption
contributions as function of energy and angle ($\mu_{exp}(E,\theta)$),
as follows: \begin{equation}
\mu_{exp}(E,\theta)=\frac{1}{\Gamma}\int_{0}^{\infty}I(E,z,\theta)\mu(E,z)dz\label{eq:mu-exp-integral}\end{equation}
where $I(E,z,\theta)$ is the radiation intensity as function of E,
z and $\theta$; $\mu(E,z)$ is the absorption spectrum contribution
at the depth z and $\Gamma$ is the normalizing factor $\int_{0}^{\infty}I(E,z,\theta)dz$.
The main difficulty to determine each $\mu(E,z)$ by solving this
equation is the initial calculation of $I(E,z,\theta)$, which depends
on how the layers structure of the film dynamically refract and reflect
the incident radiation, as a function of energy and depth. The formalism
adopted to determine this intensity and the way to extract the depth
dependence from GI-XAS spectra are described in the following section.

\subsection{Refracted and reflected amplitudes as a function of the penetration
depth, photon energy and incident angle}

Several approaches \citep{BornWolf,Henke-ADNDT93,Yun-JAP90,Mikulik-phd97,Nielsen-01,DeBoer-PRB91,Authier-2001,Baron-PhD95,Stepanov-PRB98,Lee-Haskel-PRB03,Parrat-PR54}
can be used to estimate the transmissivity of x-rays inside a material.
Those based on the dynamical diffraction theory give the most accurate
results near the critical angle of total external reflection. We deal
here with conditions near the critical energy and angle for absorption
and reflectivity resonances. Hence the method must include all dynamical
reflections and refractions conditions to determine the internal electromagnetic
wave amplitude in the samples. To fulfill these requirements we apply
an approach analogous to the recursive Parrat's reflectivity method
\citep{Parrat-PR54} to calculate the refracted and reflected amplitudes
at every depth within a film formed by $n$ layers, each one with
different chemical contributions. 

Following the definitions by \citet{Parrat-PR54}, we consider an
electromagnetic wave propagating into a material: \begin{equation}
\vec{E}=\vec{E_{0}}e^{i[\hat{n}(\vec{k}\cdot\vec{r})-\omega t]}=\vec{E_{0}}e^{i[(1-\delta)(\vec{k}\cdot\vec{r})-\omega t]}e^{-\beta(\vec{k}\cdot\vec{r})}\label{eq:campo-eletrico-raios-x}\end{equation}
 where $\delta$ and $\beta$ are the real and imaginary part of the
complex refraction index~\citep{Nielsen-01,Henke-ADNDT93,Chantler-Tables2000}
$\hat{n}=1-\delta-i\beta$. 

The continuity at each interface between the media $n$ and $n-1$
of a film with $N$ layers gives the twin equation : \begin{eqnarray}
a_{n-1}E_{n-1}+a_{n-1}^{-1}E_{n-1}^{R} & = & a_{n}^{-1}E_{n}+a_{n}E_{n}^{R}\label{eq:conservacao-eletromag}\\
\left(a_{n-1}E_{n-1}+a_{n-1}^{-1}E_{n-1}^{R}\right)f_{n-1}k_{1} & = & \left(a_{n}^{-1}E_{n}+a_{n}E_{n}^{R}\right)f_{n}k_{1}\nonumber \end{eqnarray}
where $f_{n}=(\theta_{n}^{2}-2\delta_{n}-2i\beta_{n})^{1/2}$ for
each $n$ media and $a_{n}=e^{-i\frac{\pi}{\lambda}f_{n}d_{n}}$ is
a phase factor taking into account the absorption, where $d_{n}$
is the half penetration in the media $n$. $E_{n}$ and $E_{n}^{R}$
are the total and reflected electric field amplitudes in the media
$n$. The solution of the equation \ref{eq:conservacao-eletromag}
for the reflected amplitude $R_{n-1,n}$ can be recursively determined
by \begin{equation}
R_{n-1,n}=a_{n-1}^{4}\left[\frac{F_{n-1,n}+R_{n,n+1}}{1+R_{n,n+1}F_{n-1,n}}\right]\label{eq:reflec-parrat-Rn1n}\end{equation}
where $F_{n-1,n}=\frac{f_{n-1}-f_{n}}{f_{n-1}+f_{n}}$ and $R_{n,n+1}=a_{n}^{2}(E_{n}^{R}/E_{n})$. 

The reflectivity at the interface between the air (or vacuum) and
the film is obtained after previous determination of $R_{n-1,n}$
at all others interfaces inside the film, considering that $R_{N-1,N}=0$,
as the infinitely thick substrate does not add any reflection.

The electromagnetic radiation amplitude at each depth ($z$) inside
a thin film can be determined in the same way by solving the equations
\ref{eq:conservacao-eletromag} to inform the amplitude $E_{n}$ at
each layer $n$ and depth $z$. Isolating $E_{n}$ in equation \ref{eq:conservacao-eletromag}
and using the value $R_{n-1,n}=a_{n-1}^{2}(E_{n-1}^{R}/E_{n-1})$,
one straightforwardly obtains: \begin{equation}
E_{n}=a_{n}a_{n-1}\left(\frac{1+\frac{F_{n-1,n}+R_{n,n+1}}{1+R_{n,n+1}F_{n-1,n}}}{1+R_{n,n+1}}\right)E_{n-1}\label{eq:Parrat-En-deduzido}\end{equation}

$E_{n}$ can be recursively determined from the previous knowledge
of the elements $R_{n,n+1}$ and $F_{n-1,n}$ calculated for the total
reflectivity. Consequently, the amplitude at the upper interface of
layer $n$ \textcolor{black}{is :} \begin{equation}
b_{n}=\frac{E_{n}}{a_{n}E_{0}}\left(1+R_{n,n+1}\right)\label{eq:Parrat-amp-interface}\end{equation}
with $E_{0}$ incident amplitude on the film surface ($n=0$).

\begin{figure}
\includegraphics[bb=0bp 0bp 440bp 240bp,clip,scale=0.55]{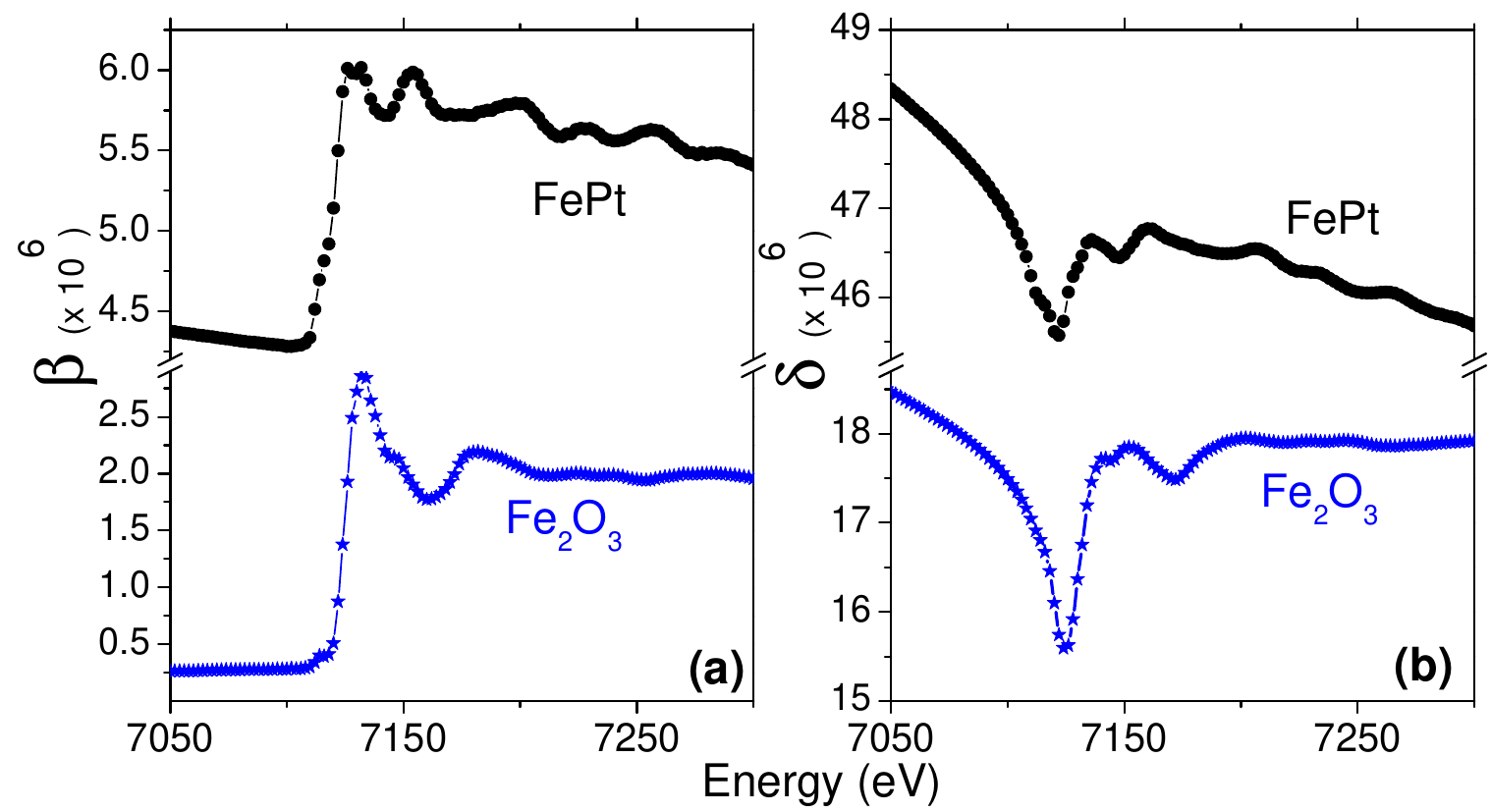}

\caption{\label{fig:ind-refrac-Fe2O3-FePt}(color online) Experimental refraction
index $\hat{n}=1-\delta-i\beta$ corresponding to FePt and $\mathrm{Fe_{2}O_{3}}$.
In (a) is the imaginary $\beta$ component, and the real $\delta$
component obtained with Kramers-Kronig transform is in (b). }

\end{figure}

The amplitude inside the layer $n$ at an arbitrary position $z_{n}$
relative to the top of the layer is then $B_{n}=b_{n}\cdot e^{-i\frac{2\pi}{\lambda}f_{n}z_{n}}$
and the radiation intensity inside this layer $n$ is $I_{n}(z_{n},\theta,E)=\left|b_{n}\cdot e^{-i\frac{2\pi}{\lambda}f_{n}z_{n}}\right|^{2}$.
The total intensity $I(z,\theta,E)$ at each depth $z$ is given by
the set of $I_{n}(z_{n},\theta,E)$ considering each thicknesses $d_{n}$
and all possible $n$. The angle and energy dependences contained
in the $f_{n}(\delta,\beta,\theta)$ complex terms are fully mathematicaly
and computationally considered, where $\beta$ and $\delta$ are the
components of the refraction index, as shown in figure \ref{fig:ind-refrac-Fe2O3-FePt}
for \textcolor{black}{FePt a}nd $\mathrm{Fe_{2}O_{3}}$~compounds.

\subsection{Depth dependence of XAS spectra}

The electromagnetic radiation intensity $I(E,z,\theta)$ described
above is used to determine the depth profile of XANES experimentally
obtained in the grazing incidence geometry. This is performed by fitting
XANES spectra for several grazing angles around the critical angle,
considering the x-ray attenuation inside the material. The result
of this process is the stratification in layers (of thickness $dz$)
of the XANES information. The XANES spectra for each depth are fitted
as a linear combination of reference spectral contributions previously
determined. As the structure of layers inside the film can change
dynamically in the fitting process, the $I(z,\theta,E)$ intensity
must be calculated at each self-consistent fitting iteration. 

To determine the depth dependence of $\mu_{exp}$, it must be found
a set of $\mu(E,z)$ data that when convoluted with $I(E,z,\theta)$
in the equation \ref{eq:mu-exp-integral} simultaneously fits $\mu_{exp}(E,\theta)$
measured for several $\theta$. This is more easily done rewriting
the equation \ref{eq:mu-exp-integral} in a discrete form considering
the sum in $z$ with steps of $\Delta z$ in depth:\begin{equation}
\mu_{exp}(E,\theta)=\frac{1}{\Gamma}\sum_{i=0}^{\infty}I(E,z_{i},\theta)\mu(E,z_{i})\label{eq:mu-exp-soma}\end{equation}
where $z_{0}=0$, $z_{\infty}=\infty$, $z_{i+1}-z_{i}=\Delta z$
and $\Gamma=\sum_{i=0}^{\infty}I(E,z_{i},\theta)$. 

It is easily seen that the proportional contribution (PC) for each
$\Delta z$ layer at depth $z_{i}$ for the signal $\mu_{exp}(E,\theta)$
is determined by $PC(z_{i})=I(E,z_{i},\theta)/\Gamma$. If the experimental
spectra can be considered as a linear combination of several independent
contributions of XANES features, $\mu_{exp}(E,\theta)$ can be considered
as a linear combination of $q$ spectral contributions (j) each one
weighed by a factor $w_{j}$:\begin{equation}
\mu_{exp}(E,\theta)=\frac{1}{\Pi}\sum_{j=1}^{q}\mu_{j}(E)\cdot w_{j}(\theta)\label{eq:mu-exp-contrib}\end{equation}
where $\Pi$ is the normalization factor $\Pi=\sum_{j=1}^{q}w_{j}(\theta)$.

Considering the equations \ref{eq:mu-exp-soma} and \ref{eq:mu-exp-contrib},
each $\mu(E,z_{i})$ can be written as $\mu(E,z_{i})=\frac{1}{\Pi}\sum_{j=1}^{q}\mu_{j}(E)\cdot w_{j}(z_{i})$
where $\Pi(z_{i})=\sum_{j=1}^{q}w_{j}(z_{i})$. Therefore:\begin{equation}
\mu_{exp}(E,\theta)=\frac{1}{\Gamma}\sum_{i=0}^{\infty}\frac{1}{\Pi}\sum_{j=1}^{q}I_{j}(E,z_{i},\theta)\cdot\mu_{j}(E)\cdot w_{j}(z_{i})\label{eq:mu-exp-sum-eq-a-ajustar}\end{equation}

The objective of determining all absorption contributions ($\mu_{j}(E)$)
and its equivalent weight ($w_{j}(z_{i})$) for each depth $z_{i}$
can be achieved by fitting the experimental spectra $\mu_{exp}(E,\theta)$
with equation \ref{eq:mu-exp-sum-eq-a-ajustar}. Is important to note
that since $\beta(E)$ for each spectral contribution at each layer
is included on both $I(E,z,\theta)$ and $\mu(E,z)$ of equation \ref{eq:mu-exp-soma},
the deconvolution of $\mu_{exp}(E,\theta)$ must be a self-consistent
procedure in terms of $\beta$. In other words, $I(E,z,\theta)$ must
be computed at every iteration of the $\mu_{exp}(E,\theta)$ fitting
procedure.

\section{Application: depth profile in a FePt magnetic thin film}

The FePt film studied here was grown by sputtering from pure targets
of elemental Fe and Pt. It was deposited on MgO(100) substrate with
a pre-deposited Pt fcc(100) buffer layer. The substrate temperature
was kept at 500\textdegree{}C to ensure a good chemical order and
perpendicular magnetic anisotropy \citep{Martins-JMMM06}. The composition
and thickness of the film was checked by Rutherford Backscattering
Spectroscopy (RBS) confirming the equiatomic ratio (51\% Fe and 49\%
Pt) and a thickness of 103 nm for the FePt layer.

\begin{figure}
\includegraphics[bb=30bp 25bp 550bp 710bp,clip,scale=0.45]{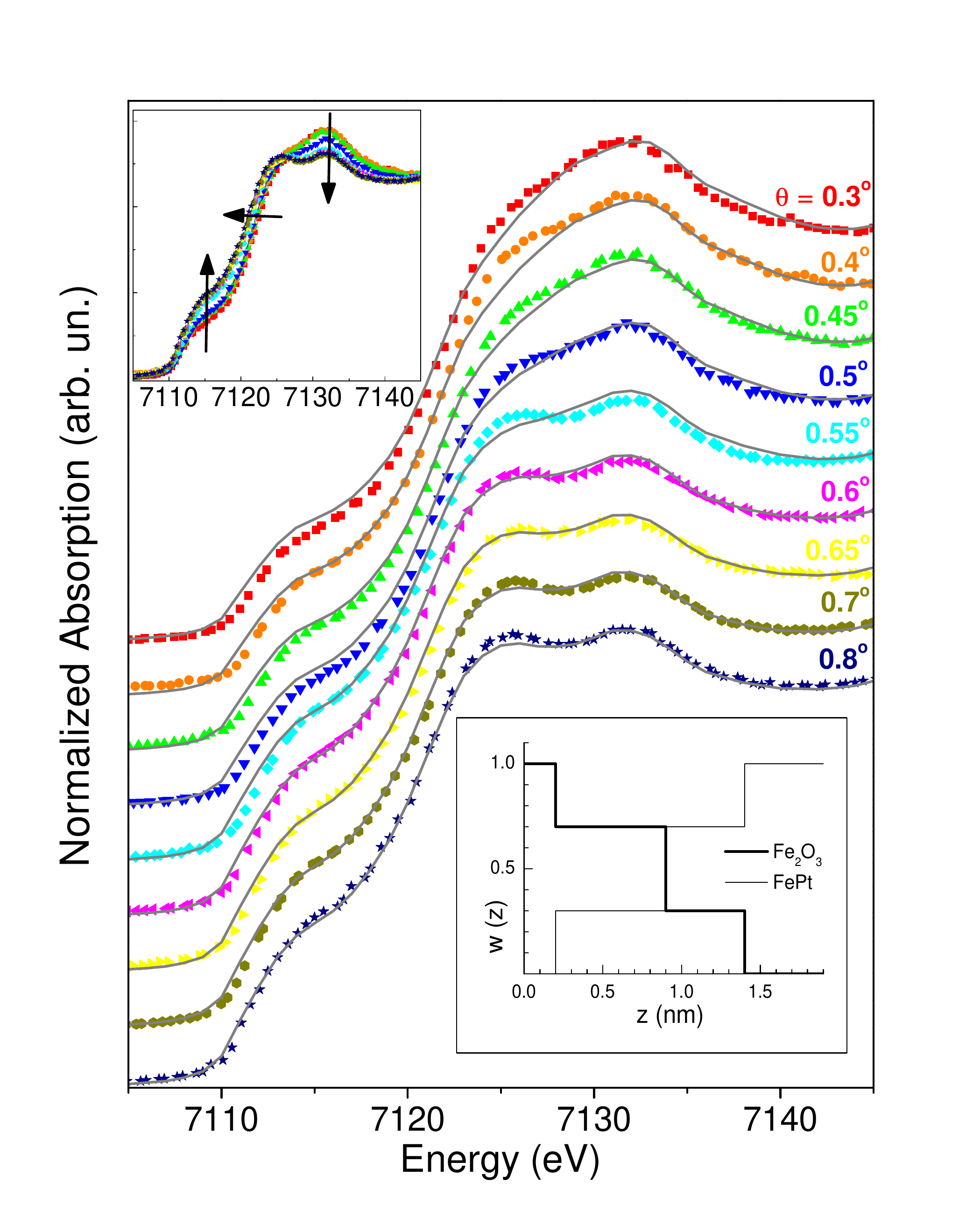}

\caption{\label{fig:ajuste-perfil-oxidoFePt}(color online) Fitted XANES experimental
data as \textcolor{black}{a} function of the energy and several grazing
angles for a FePt film. The best model to fit the experimental data
consists of a gradient between a oxidated surface with pure $\mathrm{Fe_{2}O_{3}}$
layer (0.2 nm thick), with intermediary layers of both oxide and metal
(0.7 and 0.5 nm thick), to the pure FePt metal inside the films. Upper
inset : GI-XAS measurements for increasing grazing angles showing
the dependence of the \textcolor{black}{XANES} features as \textcolor{black}{a
}function of the penetration depth, or grazing angle. The arrows correspond
to an increase in the grazing angle. Lower inset: \textcolor{black}{weight}
function w for the two components (FePt and $\mathrm{Fe_{2}O_{3}}$),
for the oxidation profile corresponding to the best fit.}

\end{figure}

When deposited at high substrate temperature (> 400\textdegree{}C)
FePt thin films without cap layer protection are easily oxidized materials
\citep{Na-FePt-oxide-IEEE2001}. The presence of a surface oxidation
is clearly observed in the GI-XAS measurements at the smallest grazing
angle presented on figure \ref{fig:ajuste-perfil-oxidoFePt}, and
the oxide is identified as $\mathrm{Fe_{2}O_{3}}$. The contribution
of the oxide layer decreases rapidly for increasing angles, indicating
that this layer is limited to a few Å. This is clearly seen by the
decreasing (increasing) feature at 7133 (7115) eV and the shift of
the spectra to lower energies presented in figure \ref{fig:ajuste-perfil-oxidoFePt}
(upper inset).

\begin{figure}
\begin{centering}
\includegraphics[bb=0bp 0bp 400bp 330bp,clip,scale=0.33]{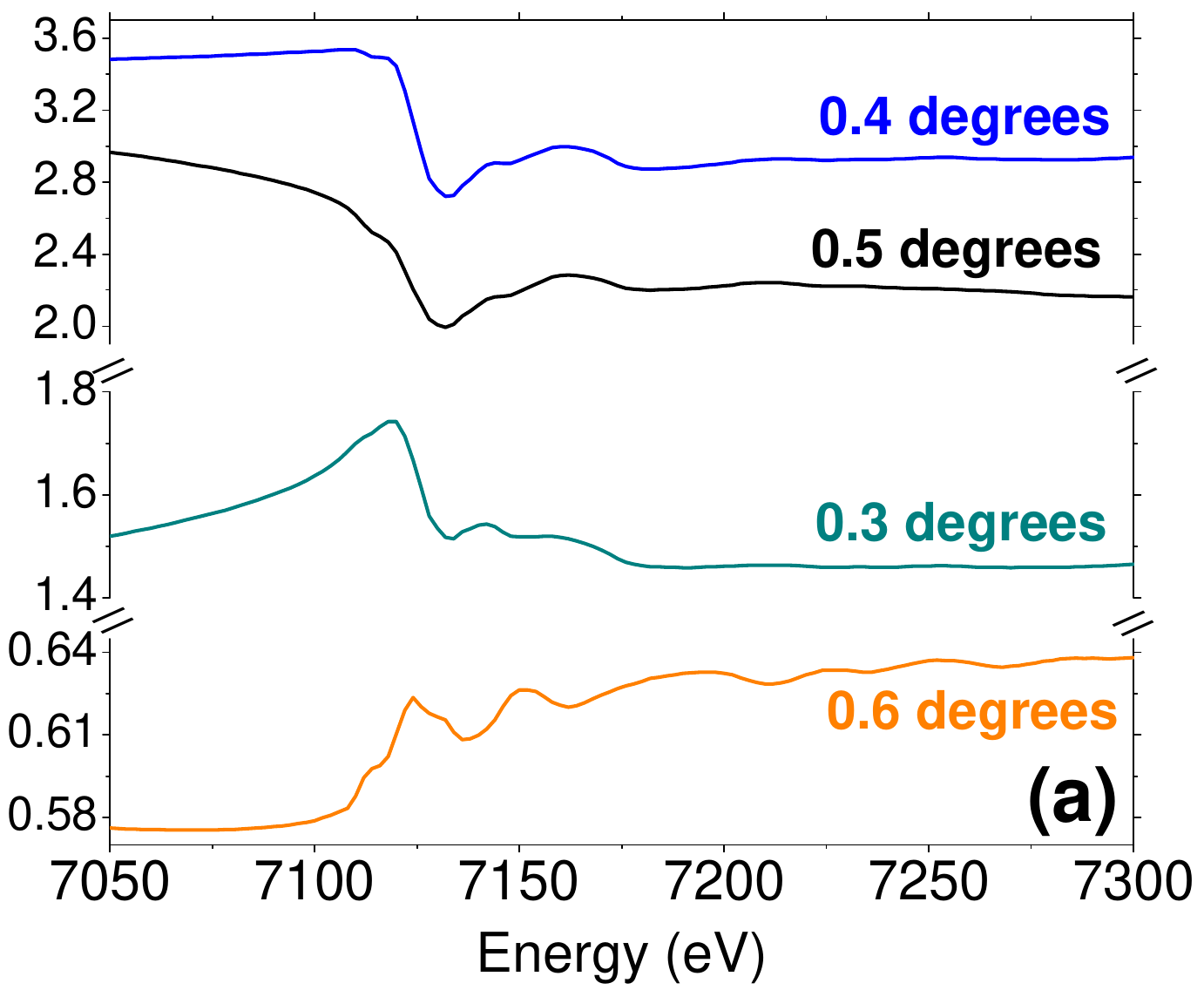}~\includegraphics[bb=0bp 0bp 380bp 370bp,clip,scale=0.3]{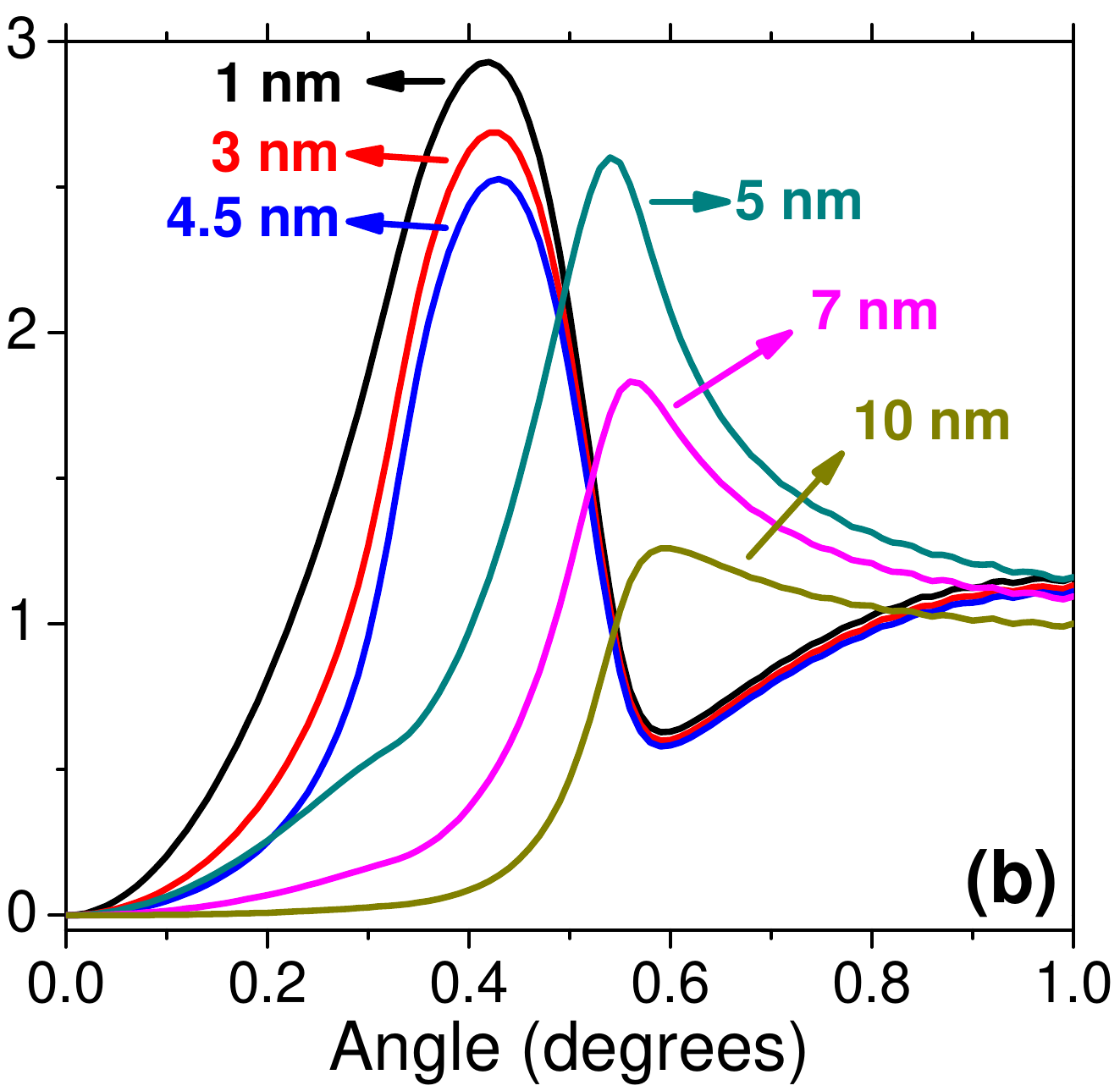}
\par\end{centering}

\caption{\label{fig:int-parrat-Fe2O3-FePt-z-theta-E}(color online) Calculated
intensity of the electromagnetic radiation inside a thin film of $\mathrm{[Fe_{2}O_{3}]_{5nm}/[FePt]_{95nm}/[Pt]_{50nm}/MgO}$.
In (a) is the intensity as function of photon energy and grazing angle,
at fixed penetration of 2 nm. In (b) is the dependence with incident
angle and the depth penetration, with fixed energy of 7130 eV. }

\end{figure}

Simple tabled or calculated $\delta$ and $\beta$ components of the
refraction index $\hat{n}=1-\delta-i\beta$ \citep{Henke-ADNDT93,Chantler-Tables2000}
cannot be used to calculate $I(E,z,\theta)$, given that it would
include approximations not valid when the spectral features near a
critical energy (absorption edge) are the desired information. The
imaginary $\beta$ component of the experimental refraction index
was obtained for reference FePt and $\mathrm{Fe_{2}O_{3}}$ samples
using their absorption spectra scaled to tabled absolute values far
from the absorption edges \citep{Henke-ADNDT93,Chantler-Tables2000},
as shown in figure \ref{fig:ind-refrac-Fe2O3-FePt}a. Kramers-Kronig
transforms\citep{Ohta-AS88,Bertie-CJC92,PETERSON1973,HOYT1984,King-JOSAB02,King2006,Cross-RSI99}
were used to determine the correspondent real $\delta$ component
(fig. \ref{fig:ind-refrac-Fe2O3-FePt}b). \textcolor{black}{Figure
\ref{fig:int-parrat-Fe2O3-FePt-z-theta-E} shows the simulated intensity
$I(E,z,\theta)$ for a model thick FePt film with a flat 5 nm $\mathrm{Fe_{2}O_{3}}$
layer on its surface, using equation \ref{eq:Parrat-amp-interface}
and the refraction index data for each layer presented in figure \ref{fig:ind-refrac-Fe2O3-FePt}}.
The XANES structures must definitively be taken into account when
determining the experimental refraction index in order to include
all energy/angle dependences in $I(E,\theta)$, as is ratified by
the strong non-linear dependence on $I(E)$ for different grazing
angles shown in fig. \ref{fig:int-parrat-Fe2O3-FePt-z-theta-E}a.
The intensity $I(\theta)$ at $E$= 7130eV for representative $z$
values is shown in fig. \ref{fig:int-parrat-Fe2O3-FePt-z-theta-E}b,
which exemplifies multiple reflection interference effects near the
critical angle for each penetration. The interference condition for
the first $\mathrm{Fe_{2}O_{3}}$ layer (5 nm) is drastically different
from the resonances for the deeper layers, where the contributions
arise essentially from the FePt alloy. These simulations emphasizes
the strong need to consider all dynamical reflections at the interfaces
to accurately calculate $I(E,z,\theta)$.

On the other hand it is worth noting that although corrections due
to fluorescence self-absorption effects might be important in some
cases, these effects are not significant for the angular range and
penetration depth discussed here (<1\% in the final error bars). 

Equation \ref{eq:mu-exp-sum-eq-a-ajustar}, considering the iteratively
determined $I(E,z,\theta)$, was used to simultaneously fit the experimental
XANES spectra taken at several grazing angles (fig. \ref{fig:ajuste-perfil-oxidoFePt}).
Reference XANES spectra of $\mathrm{Fe_{2}O_{3}}$ and FePt were used
in the fit. 

The several experimental spectra simultaneously fitted, shown in figure
\ref{fig:ajuste-perfil-oxidoFePt}, enable us to determine the complex
\textcolor{black}{layers} structure in the depth profile, beyond a
simply oxidized thickness determination. Different models of layers
structure were considered for the oxidized FePt surface. Although
a flat top oxide layer is well-suited to illustrate the general behavior,
it is by far not the right solution to fit our data. The depth profile
analysis shows clearly that the oxide not only covers the FePt film
but penetrates beneath the film giving rise to a fractionated buried
layer composed of the oxide and FePt. The best fit model\textcolor{black}{
turns }out to be a gradient between a thin oxidized surface with pure
$\mathrm{Fe_{2}O_{3}}$ layer and intermediate layers of both oxide
and metal alloy down to 1.4 nm from the surface. The top 100\% oxide
layer is 0.2 nm thick, followed by two mixed layers with 70\% and
30\% of $\mathrm{Fe_{2}O_{3}}$ and 0.7 and 0.5 nm thick, respectively.
The weight function w for each component is shown in the lower inset
in figure \ref{fig:ajuste-perfil-oxidoFePt}. 

It has been reported \citep{Na-FePt-oxide-IEEE2001} that for FePt
films an Fe oxide layer would form on the surface due to Fe migration
to the oxide/metal interface during the growth at high temperatures.
As a result, there might be a composition variation with increasing
film depth. In the metallic layer just below the oxide, Fe content
should be lower than that of as-deposited film while Pt content should
be higher. Our results confirm a compositional variation over the
film depth, but supports a more complex picture. As known from literature,
sputtered metallic films have some tendency to form pillars. We interpret
the gradient as resulting from the decoration of these pillars by
the oxide that fills the empty space between pillars and oxidizes
the very interfacial Fe atoms, rather than a continuous rough surface. 

We should finally include an additional remark about the resolution
of the depth profiles. Due to the exponential decay of the radiation
intensity inside the film, the depth probed and final resolution of
the method are intrinsically correlated and strongly dependent to
the contrast between the refraction index of each layer material in
the film. For instance, in the $\mathrm{Fe_{2}O_{3}}$/FePt case the
refraction index of $\mathrm{Fe_{2}O_{3}}$ is factor three smaller
than for the FePt material. In this case the profile variation is
confined within 2 nm near the surface and the depth resolution is
of order of one angstrom. If the compositional gradient were deeper
into the film, the profile resolution would be lower for the internal
layers.

\section{Conclusion}

Scattering and absorption phenomena are intrinsically intercorrelated
when grazing incidence reflection and refraction are combined to x-ray
absorption spectroscopy. The approach of GI-XAS presented in this
article fully considers both scattering and absorption formalism for
depth profiling the atomic, electronic, chemical or magnetic local
structures in thin films with nanometric resolution. This formalism,
not facing intrinsic limitations or approximations, can be applied
to deconvolve the depth dependencies of not only XANES information
as exemplified here, but also XRF (x-ray fluorescence) and XMCD (x-ray
magnetic circular dichroism) signals in the fluorescence or reflectivity
channels from thin films and multilayers.

\begin{acknowledgments}
This work is partially supported by LNLS/ABTLuS/MCT. NMSN acknowledges
the grants from CNPq and CAPES. 
\end{acknowledgments}

\bibliographystyle{apsrev}

\end{document}